\documentclass[useAMS,usenatbib,referee]{mn2e}
\usepackage{graphicx}
\usepackage{graphics}
\usepackage{multicol}
\usepackage{subfig}

\title[The torque reversals and pulse profile of 4U~1626--67]{Torque reversals and pulse profile of the pulsar 4U~1626--67}

\author[A. Beri et al.]
{Aru~Beri,$^1$
Chetana~Jain,$^2$ 
 Biswajit~Paul,$^3$ Harsha~Raichur,$^3$ \\
$^1$, Department of Physics, Indian Institute of Technology Ropar, Nangal Road, Rupnagar, Punjab, 140001 India\\
$^2$, Hans Raj College, University of Delhi, Delhi-110007, India \\
$^3$, Raman Research Institute, Sadashivnagar, C. V. Raman Avenue, Bangalore-560 080, India.\\}

\begin{document}

\pagerange{\pageref{firstpage}--\pageref{lastpage}} 
\maketitle
\label{firstpage}

\begin{abstract}
We review the pulse profile evolution of the unique accretion powered X-ray pulsar 4U~1626--67 over the last 40 years since its discovery. This pulsar showed two distinct eras of steady spin-up separated by a steady spin-down episode for about 18 years. In the present work, using data from different observatories active during each phase of spin-up and spin-down we establish a clear correlation between the accretion torque acting on this pulsar and its pulse profile. The energy resolved pulse profiles are identical in both the spin-up eras and quite different in the spin-down era, especially in the low energy band. This correlation, along with the already known feature of strong Quasi Periodic Oscillations (QPO) that was present only in the spin-down era, clearly establish two different accretion modes onto the neutron star which produce different pulse profiles and only one of which produces the QPOs. 

\end{abstract}


\section{Introduction}

One of the important manifestations of the interaction between the accretion disc around a neutron star and its magnetic field is the accretion torque on the neutron star, which sometimes results into transitions between spin-up and spin-down of the star.  In the standard model of accretion onto magnetised neutron stars, the accretion torque is related to the mass accretion rate onto the neutron star and hence the bolometric X-ray luminosity of the source. Thus it is expected that the spin transition takes place when the source luminosity decreases below a critical low luminosity which in turn depends on the spin and magnetic moment of the neutron star {\citep{Ghosh79}}. Transient Be/X-ray binary pulsars and persistent X-ray pulsars which show large X-ray flux variation are therefore ideal sources to verify the standard model of accretion onto high magnetic field neutron stars. Many of these sources have shown transitions between spin-up to spin-down and vice versa {\citep{Bildsten97}}.  However the torque onto the neutron star and the X-ray luminosity (and therefore mass accretion rate) are not well correlated in either the persistent pulsars with supergiant companions or the low mass X-ray binary pulsars {\citep{Bildsten97}}. Hence applicability of the standard model of accretion onto the magnetised neutron star is questionable for a large fraction of the X-ray pulsars.

4U~1626--67 is a persistent X-ray source, in which episodes of steady spin-up and spin-down of the pulsar have been observed, and hence is a suitable candidate for a critical examination of the standard model. Unlike most other accreting pulsars both transient and persistent \citep{Bildsten97}, transfer of angular momentum is relatively smooth in 4U~1626--67. This source was discovered with the Uhuru satellite \citep{Giacconi72} and pulsations with a period of $\sim$ 7.7 s was discovered with \emph{SAS-3} \citep{Rappaport77} and has been observed many times since then. Since its discovery, the pulsar was observed to be spinning up until 1990. One feature that stands out the most in this pulsar is the smooth spin change and occurrence of two torque reversals in a time frame of about 40 years.

During the first spin-up phase (between 1977-1990), observations made with several observatories, like \textit{SAS-3, Einstein, Ginga and EXOSAT}, established a steady spin-up rate of $\sim 8.54 \times 10^{-13}$ Hz s$^{-1}$ \citep{Joss78, Elsner83, Levine88, Shinoda90}. X-ray luminosity was estimated to be about 1 $\times$ 10$^{37}$ erg s$^{-1}$ \citep{White83}. {\cite{Chakrabarty97}} noticed a gradual decrease in X-ray flux during this phase.   The pulse profiles analyzed using the data of \emph{HEAO} \citep{Pravdo79}, \emph{TENMA} \citep{Kii86} and \emph{EXOSAT} \citep{Levine88} showed strong energy dependence.  After the torque reversal in 1990, the neutron star in 4U~1626--67 started spinning down at a rate similar in magnitude to the earlier spin-up rate {\citep{Chakrabarty97}}. After about 18 years of steady spin down, the source again underwent a torque reversal in the beginning of 2008 {\citep{Camero10, Jain10}}. After both the torque reversals, changes in the pulse profile were recorded \citep{Krauss07,Jain08,Jain10}. {\citet{Krauss07}} also reported a decrease in X-ray flux during this phase. The X-ray flux increased by more than a factor of two during the second torque reversal \citep{Camero10,Jain10}. During the first spin-up era, weak and broad Quasi Periodic Oscillations (QPO) at 40 mHz were reported from the \emph{Ginga} observations {\citep{Shinoda90}}. However, observations made with \emph{Beppo-SAX}\citep{Owens97}, \emph{Rossi X-ray Timing Explorer (RXTE)} \citep{Kommers98,Chakrabarty98} in the spin-down era show strong QPOs at around 48 mHz with a slow frequency evolution with time {\citep{Kaur08}}. These 48 mHz QPOs disappeared after the second torque reversal {\citep{Jain10}}.

The X-ray spectrum of 4U~1626--67 shows two continuum components: a hard power law and a black body. It also shows strong emission lines from oxygen and neon {\citep{Angelini95, Schulz01, Krauss07, Camero12}} the strength of which vary with time. In the first spin-up era, the spectrum had a power-law photon index of $\sim 1.5$ and a blackbody temperature of $\sim 0.6$ keV \citep{Pravdo79, Kii86, Angelini95, Vaughan97}. In the spin-down era, the energy spectrum became relatively harder with a power law index of $\sim$ 0.4-0.6 and the blackbody temperature decreased to $\sim$ 0.3 keV \citep{Angelini95, Owens97, Vaughan97, Orlandini98, Yi99}.  The X-ray spectrum measured with the Proportional Counter Array (PCA) onboard {\it RXTE} {\citep{Jain10}} and Suzaku {\citep{Camero12}} in the second spin-up phase showed a reversal with a photon index in the range of 0.8-1.0 and a higher black body temperature of about 0.5-0.6 keV. The X-ray spectrum in the second spin-up era shows the temperature quite similar to that in the first spin-up era, but the photon index did not return to the same value as that observed during \emph{Phase I} {\citep{Angelini95, Orlandini98, Schulz01, Krauss07, Jain10, Camero10, Camero12}. The detection of cyclotron line feature in its spectrum at around 37~keV with several observatories during spin-up and down phases \citep{Pravdo78, Orlandini98, Heindl99, Coburn02, Camero12, Iwakiri12} gives a measure of the surface magnetic field of $\sim $$3\times10^{12}$~Gauss.

In this work, we study the evolution of the pulse profile of 4U~1626--67 during the three phases of accretion torque using data from different observatories.  Hereafter, the spin-up era between 1977 - 1990 will be referred to as \emph{Phase I}, spin down between 1990 - 2008 as \emph{Phase II} and the current era as \emph{Phase III}. The observations made with different observatories in each phase were used to bring out the similarities and/or dissimilarities of the three phases. In the light of our results, we discuss the possible nature and behavior of torque transfer in this system.

\section{Observations, Analysis and Results}

\subsection{Representative observations in each phase}

For creating the pulse profiles we analyzed one representative long observation in each of the three phases for which archival data is available with sub-second time resolution. The longest observation
in \emph{Phase-I} was made with \emph{EXOSAT} in 1986 \citep{Levine88}. Unfortunately archival data of
\emph{EXOSAT} is available only in the form of spectra and light curve with 10 s resolution which is not suitable for
the present analysis. Therefore, for \emph{Phase I} we analyzed  the longest observation made with the Large
Area Counter (LAC) onboard \emph{Ginga} observatory \citep{Makino87} while for the other two phases the
longest observations made with the Proportional Counter Array (PCA) of the \emph{RXTE} observatory were
used \citep{Bradt93}. Details of these observations are given in Table-1. \\

The \emph{Ginga}-LAC covers the energy range of 1.5--37~keV with a photon collection area of 4000 cm$^{2}$.  Time series from the LAC data was extracted using the tools \textsc{lacqrdfits} and \textsc{timinfilfits}. Background subtraction from the light curve was done using the Hayashida method\footnote{(http://darts.jaxa.jp/astro/ginga)}.  The {\emph{RXTE-PCA}} consisted of an array of five collimated xenon/methane multi anode proportional counter units (PCU), with a maximum photon collection area of 6500 cm$^{2}$ \citep{Jahoda96, Jahoda06}.  PCA data collected in the Good Xenon mode, was used to generate the source light curves in the energy band of 2$-$30~keV. Background light curves in the same energy band were estimated using the tool \textsc{runpcabackest} assuming a faint source model as suggested by \emph{RXTE} GOF\footnote{(http://heasarc.gsfc.nasa.gov/docs/xte/pca$_{-}$news.html)}. Thereafter, the photon arrival times in the background subtracted light curves were corrected to the solar system barycenter.

\begin{table}
\centering
\caption{Log of observations of 4U 1626-67 used in figures 1-4}
\begin{tabular}{@{}lccc@{}}

\hline
\hline

Observatory             &  Year & Observation ID & Total Exposure (ks)  \\
\hline
\emph{Ginga}            & 1988      & 880712600             & 17  \\
\emph{RXTE}             & 1996      & 10101-01-01-00        & 73  \\
\emph{RXTE}             & 2010      & 95313-01-01-03        & 32    \\

\hline
\end{tabular}
\end{table}

\subsubsection{Light curves}

Figure \ref{lightcurve} shows the barycenter corrected and background subtracted 2-30~keV light curves in each \emph{Phase}.  The plotted light curves have been binned with a bin size of about 10 pulses (77 s).  It is obvious from Figure \ref{lightcurve} that flaring events are observed with \emph{Ginga} and \emph{RXTE} detectors during both the spin-up era. Similar flaring events in both the X-ray and optical data have also been reported earlier {\citep{Joss78, McClintock80, Li80, Schulz11}. On the other hand, \emph{Phase II} light curve shows lesser variations (also reported by \cite{Krauss07} using \emph{XMM-Newton(PN)} data).

\begin{figure}
\includegraphics[height=6.5in, width=5.0in, angle=-90]{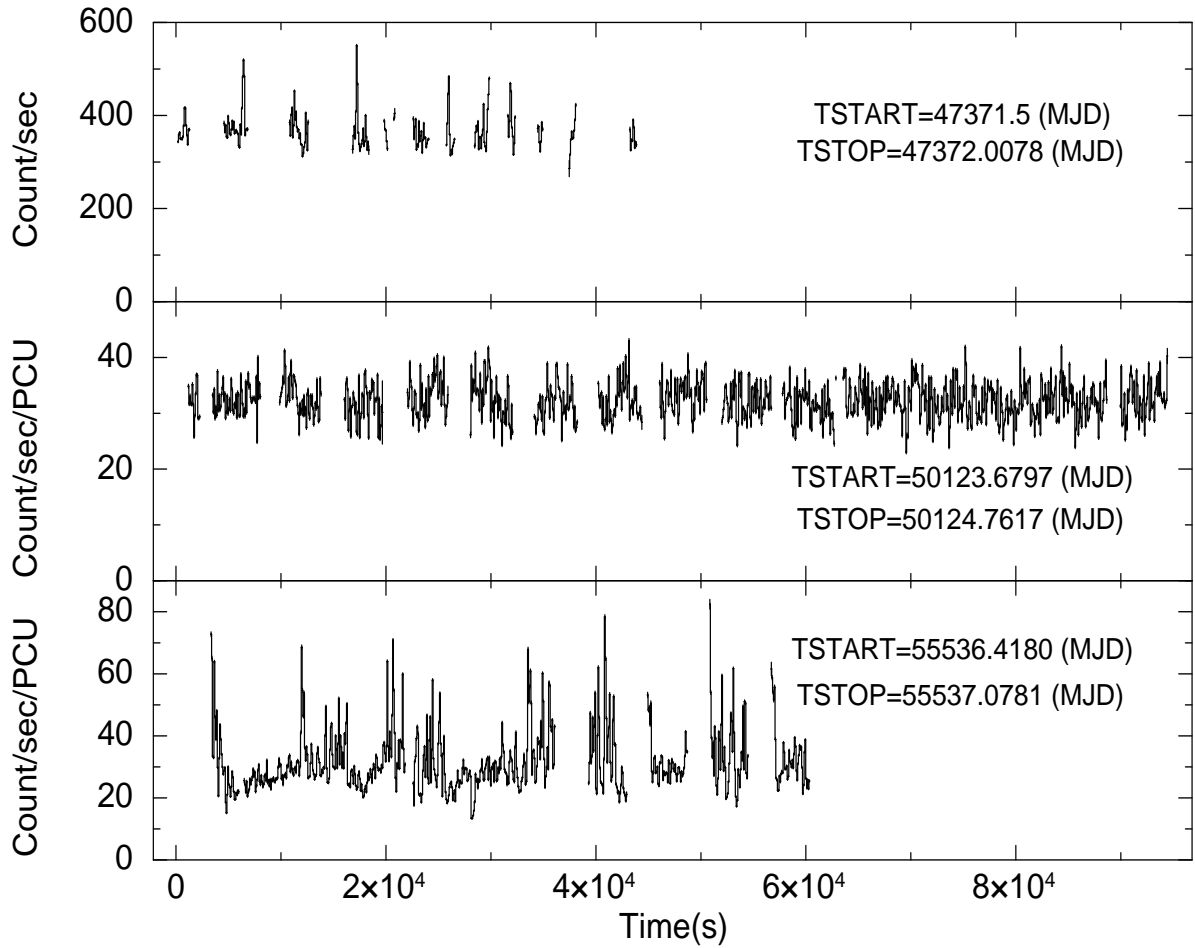}
\caption{Background subtracted, barycenter corrected 
light curves of 4U~1626--67, obtained from 
\emph{Ginga} observations, during \emph{Phase I} (top panel); 
and from $RXTE$-PCA observations during \emph{Phase II} (middle panel) 
and \emph{Phase III} (bottom panel). X-axis show time (in seconds) 
since the beginning of the respective observation. Refer to Table-1 for
details.}
\label{lightcurve}
\end{figure}

\subsubsection{Quasi Periodic Oscillations}

In Figure \ref{QPO} we have shown the Power Density Spectra (PDS) of 4U~1626--67 in the three phases. We have divided the light curves into stretches of 2048 seconds. PDS from all the segments were averaged to produce the final PDS and were normalized such that their integral gives squared rms fractional variability and the white noise level was subtracted. A strong QPO feature at $\sim$48 mHz is clearly seen in PDS of \emph{Phase II} data. Presence of strong QPOs in all the Phase II observations of this
source were reported earlier \citep{Kaur08}. Absence of the QPOs in \emph{Phase III} and a change in the
shape of the PDS from \emph{Phase II} to \emph{Phase III} is already known \citep{Jain10}. The PDS
characteristics in \emph{Phase I} are also different from that in \emph{Phase II}. In only two of the \emph{Phase I} observations QPOs have been observed, with quite different parameters. \citet{Kaur08} reported
the presence of a very narrow QPO in the \emph{EXOSAT-ME} light curve at around 35 mHz, while a weak
QPO feature at 40 mHz was seen mainly in the 14-18~keV range of this \emph{Ginga-LAC} observation \citep{Shinoda90}, which we have reconfirmed.

\begin{figure}
\includegraphics[height=2.0in, width=2.0in, angle=0]{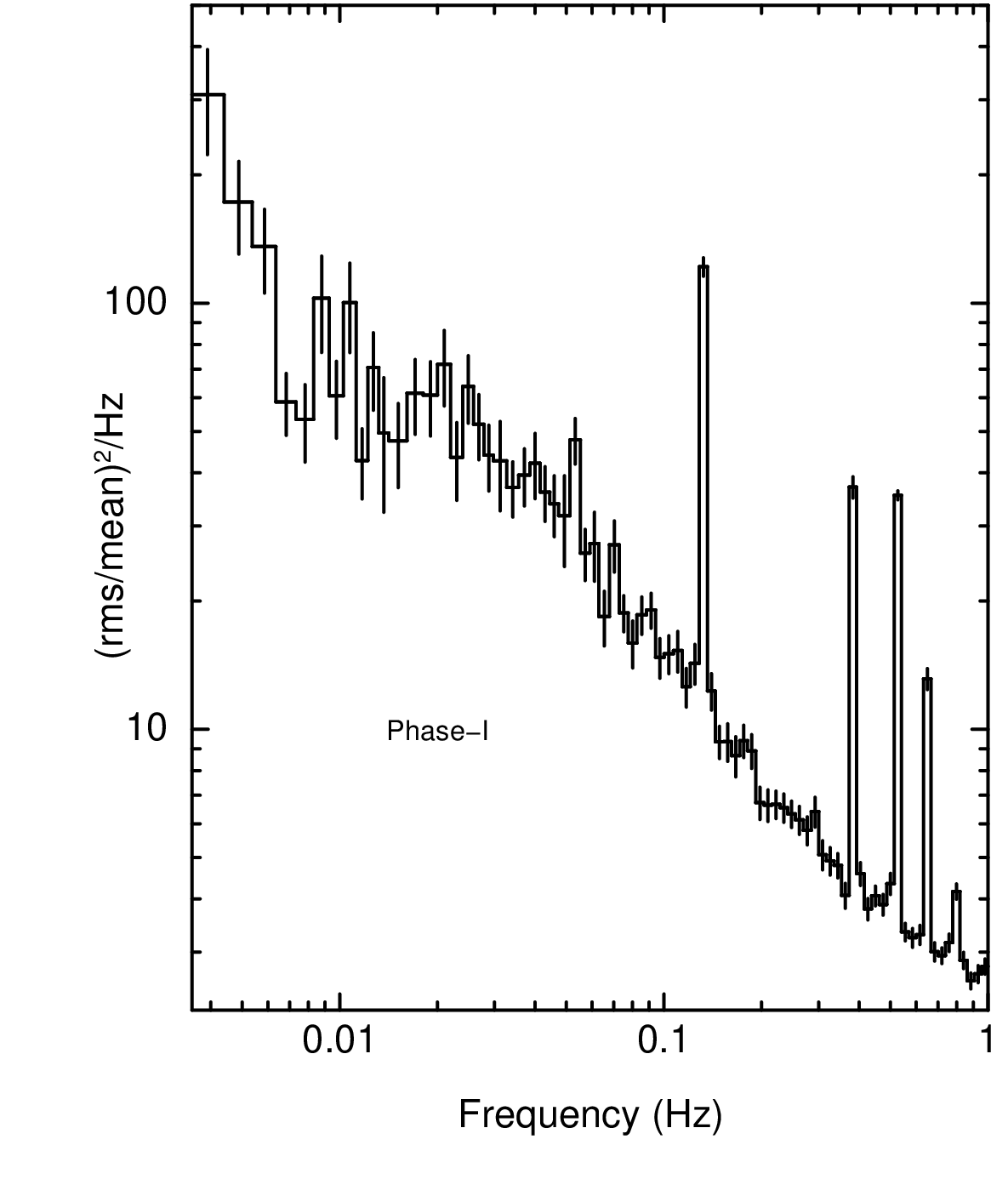}
\includegraphics[height=2.0in, width=2.0in, angle=0]{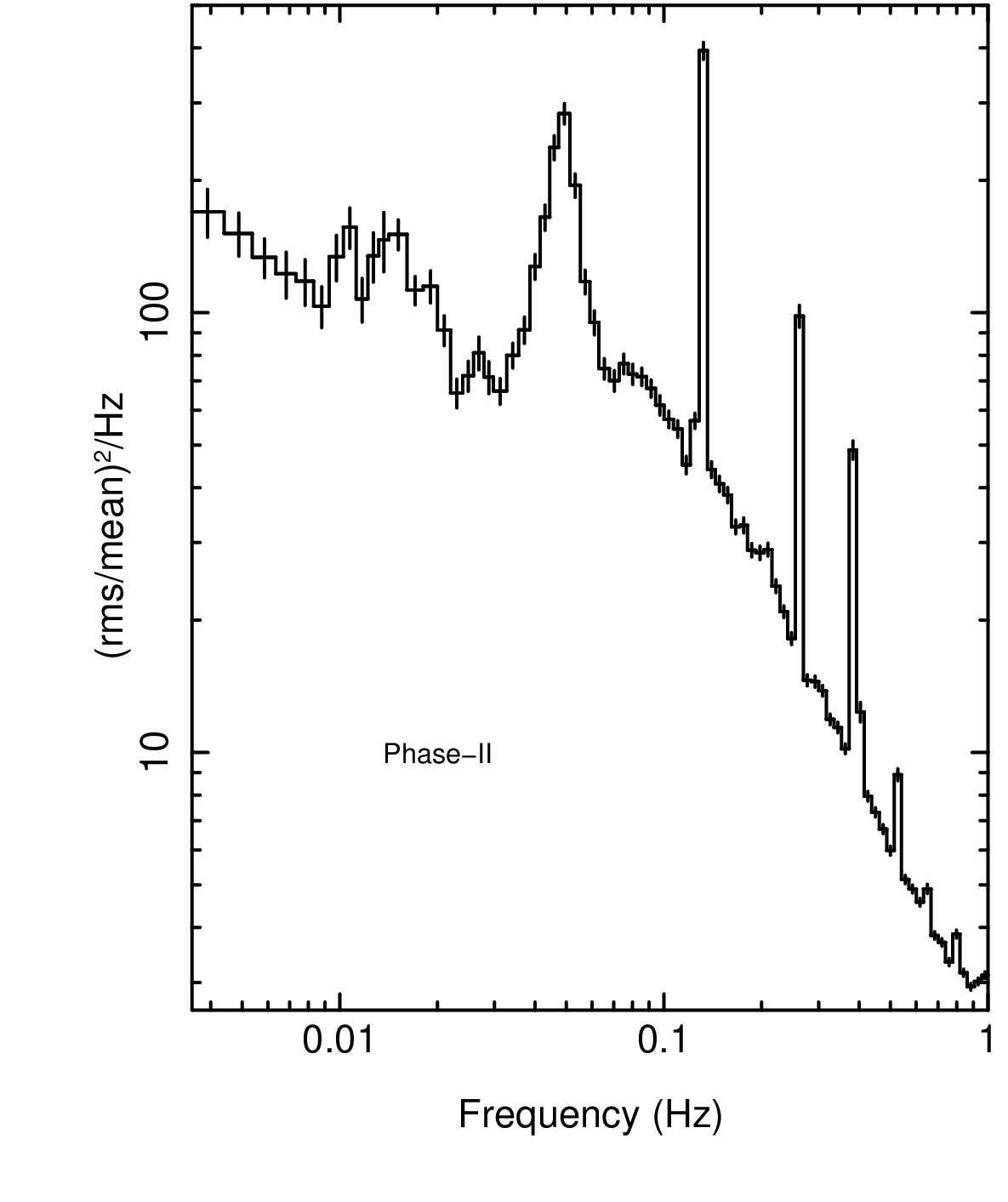}
\includegraphics[height=2.0in, width=2.0in, angle=0]{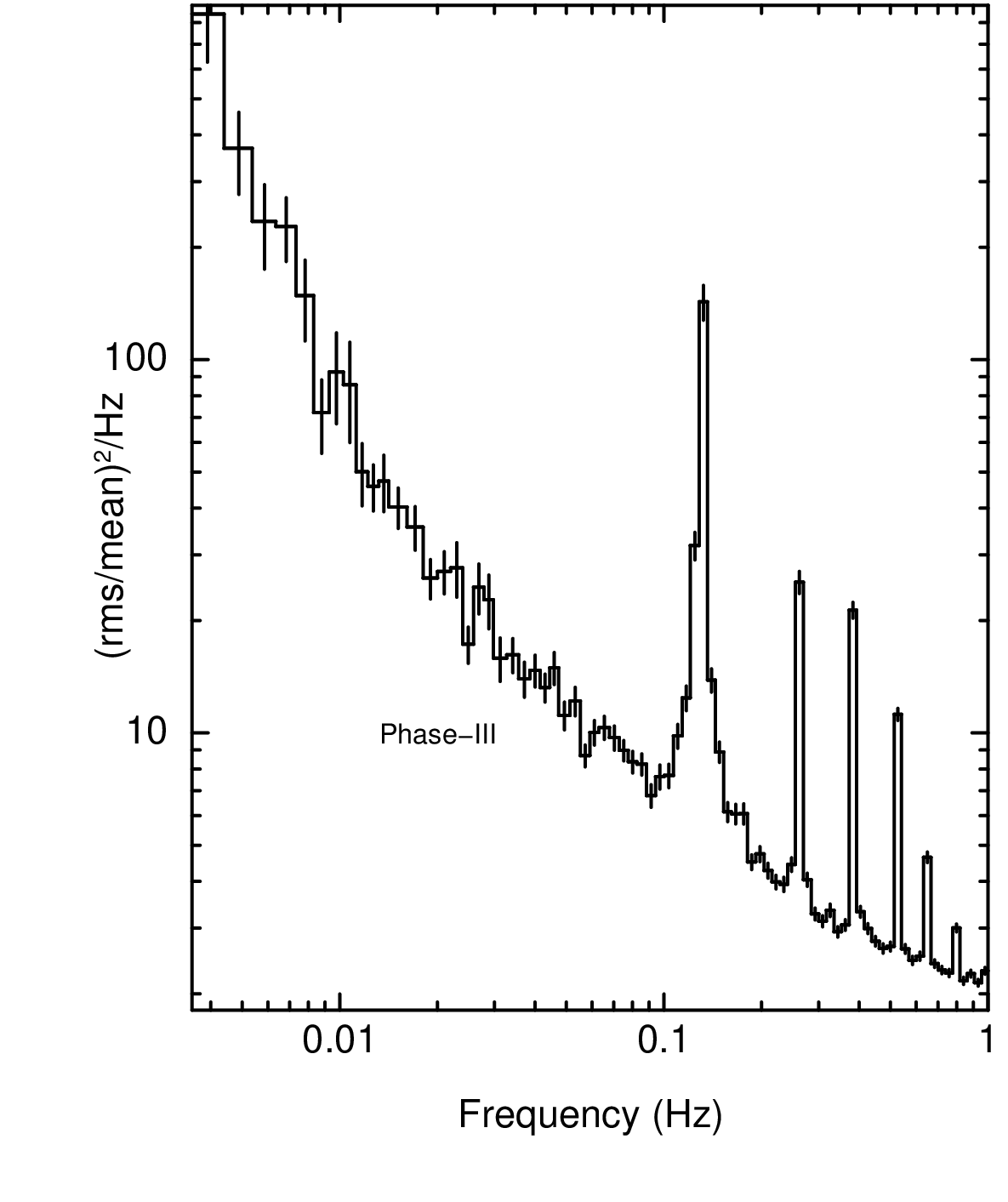}
\caption{Power spectrum of 4U~1626--67, during the three
phase of pulser spin. From left to right, the power density
spectrum of first spin-up, spin-down and second spin-up era
are shown, respectively. }
\label{QPO}
\end{figure}

\subsubsection{Broad band pulse profiles}

For each of the selected observations in the three \emph{Phases} we obtained local spin period $P_{spin}$ using the epoch folding $\chi ^{2}$ maximization technique. The pulsar had a $P_{spin}$ of 7.663 s, 7.6668 s and 7.6772 s during these three observations in \emph{Phase-I}, \emph{Phase-II} and \emph{Phase-III} respectively. The 2-30~keV light curves were folded at the respective spin periods. Figure \ref{Pulse_profile} shows the average 2-30~keV pulse profiles, each having 64 phase bins, during the three \emph{Phases}.  The pulse profiles have been shifted in pulse phase such that the minimum appears at pulse phase 1.0. 

\begin{figure}
\centering
\includegraphics[scale=0.4, angle=0]{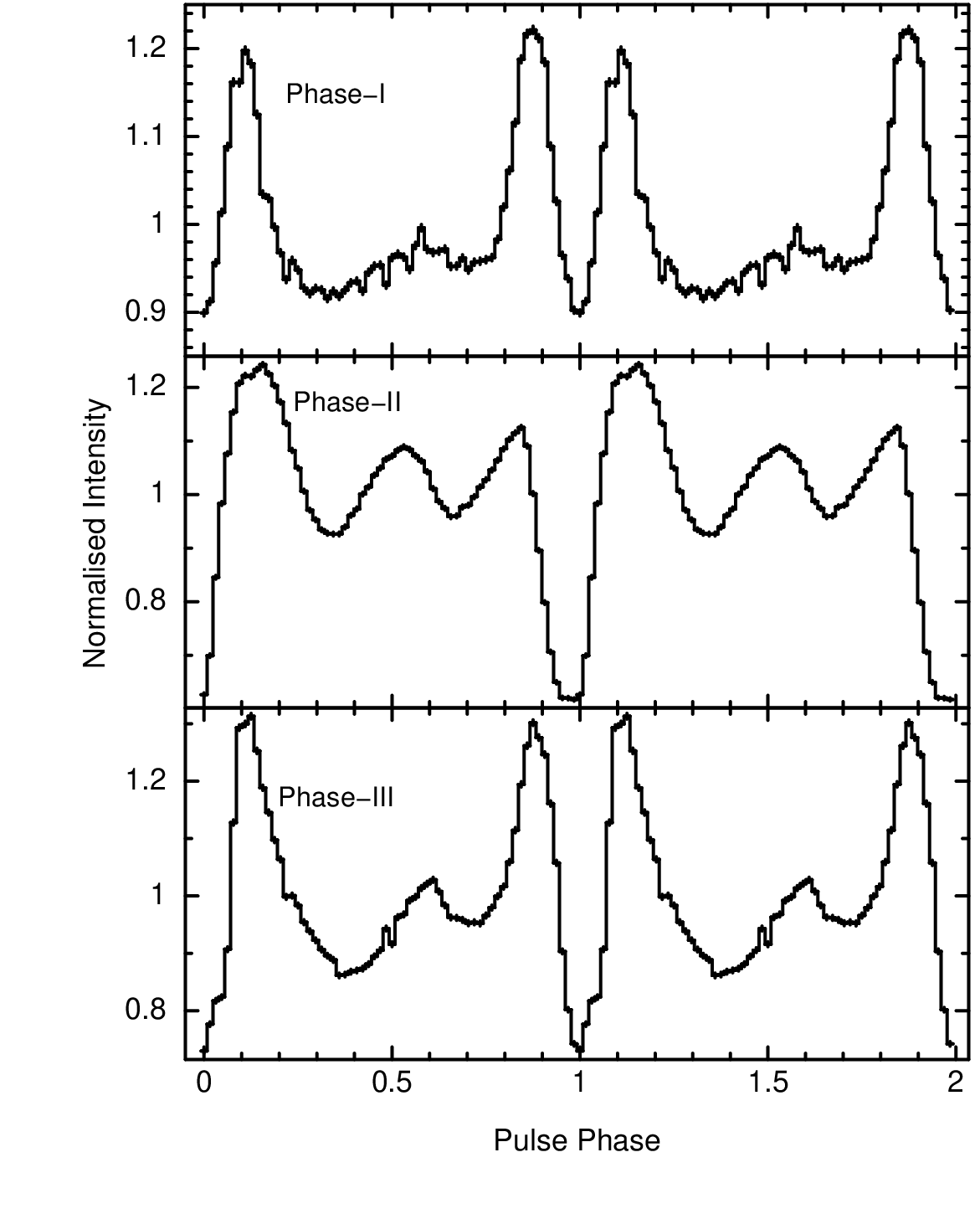}
\caption{2-30 keV pulse profile of 4U~1626--67, 
each divided into 64 phase bins. \emph{Top}: \emph{Ginga} data - 
\emph{Phase I}, \emph{Middle}: \emph{RXTE} data - 
\emph{Phase II}, \emph{Bottom}: \emph{RXTE} data - 
\emph{Phase III}. The pulse profiles have been shifted in 
pulse phase such that the minimum appears at pulse phase 1.0 }
\label{Pulse_profile}
\end{figure}

The broad band pulse profile, as detected using the \emph{Ginga} and \emph{RXTE} data has changed significantly with time. The pulse profile during \emph{Phase I} shows two peaks, with the amplitude of the first peak being slightly more than that of the second peak. The two peaks are separated by a narrow dip and the pulse profile also shows a broad minimum spanning about 0.5 in pulse phase.  The \emph{Phase II} pulse profile, on the other hand, does not show the two peaks shown in \emph{Phase I}. The $``$broad minimum$"$ feature has also disappeared and the dip is broader and deeper.  Two, shallow dips have also appeared at phase 0.4 and 0.7 after the main dip which become broader in \emph{Phase III} pulse profile. The broad band pulse profile in \emph{Phase III} has features sharper than those in \emph{Phase II} and there is a clear indication of the double peak feature. The pulse fraction $({I_{max} -I_{min}})/{I_{max}}$ changed from $\sim$26\% during \emph{Phase I}, to $\sim$50\% during \emph{Phase II} and $\sim$44\% during \emph{Phase III}. Here we note that the pulse profile of 4U~1626--67 was measured many times with various observatories in the three phases and in each of the phases, the reported pulse profiles are similar to the profiles shown in Figure \ref{Pulse_profile} (\emph{Phase I}: \cite{Rappaport77, Joss78, Pravdo79, Elsner83, Kii86, Levine88, Mihara95}; \emph{Phase II}: \cite{Angelini95, Orlandini98, Krauss07,   Jain08, Jain10,  Iwakiri12}; \emph{Phase III}: \cite{Jain10}).

\subsubsection{Energy resolved pulse profiles}

The pulse profile of 4U~1626--67 is known to have strong energy dependence, with a complex pulse shape in the low energy band and a simple broad pulse profile in the high energy band.  Hence, we also compare the pulse profiles in smaller energy bands.  The energy resolved pulse profiles in energy bands of 2-5~keV, 5-8~keV, 8-12~keV, 12-20~keV and 20-30~keV are shown in Figure \ref{ene-dependent-profile}.  The pulse profiles have many features in the low energy band and at eneriges above 20 keV they show a single, broad peak. 

The most remarkable feature in Figure 4, is the difference in the energy resolved pulse profiles of the two spin-up eras and the spin-down era. The pulse profiles of \emph{Phase I} and \emph{Phase III} across the energy band are quite similar except in the narrow range of 8-12 keV where the pulse profile of \emph{Phase-III} bears more resemblance to the respective \emph{Phase-II} pulse profile.  The feature that stands out the most is the disappearance of the sharp double peaked pulse profile during spin-down era of \emph{Phase II} and its reappearance in \emph{Phase III}. During the two spin-up era, with increasing energy the dips around pulse phase 0.1 broadens and the double peaks disappear.  Also the first of the double peaks in \emph{Phase I} is of greater amplitude than the second peak whereas in \emph{Phase III}, the second peak has greater amplitude than the first peak.  In the pulse profile during the spin-down era, there occurs a deep narrow \emph{dip} at low energies which broadens at higher energies.

\begin{figure*}
\includegraphics[height=4.0in, width=2.0in, angle=0]{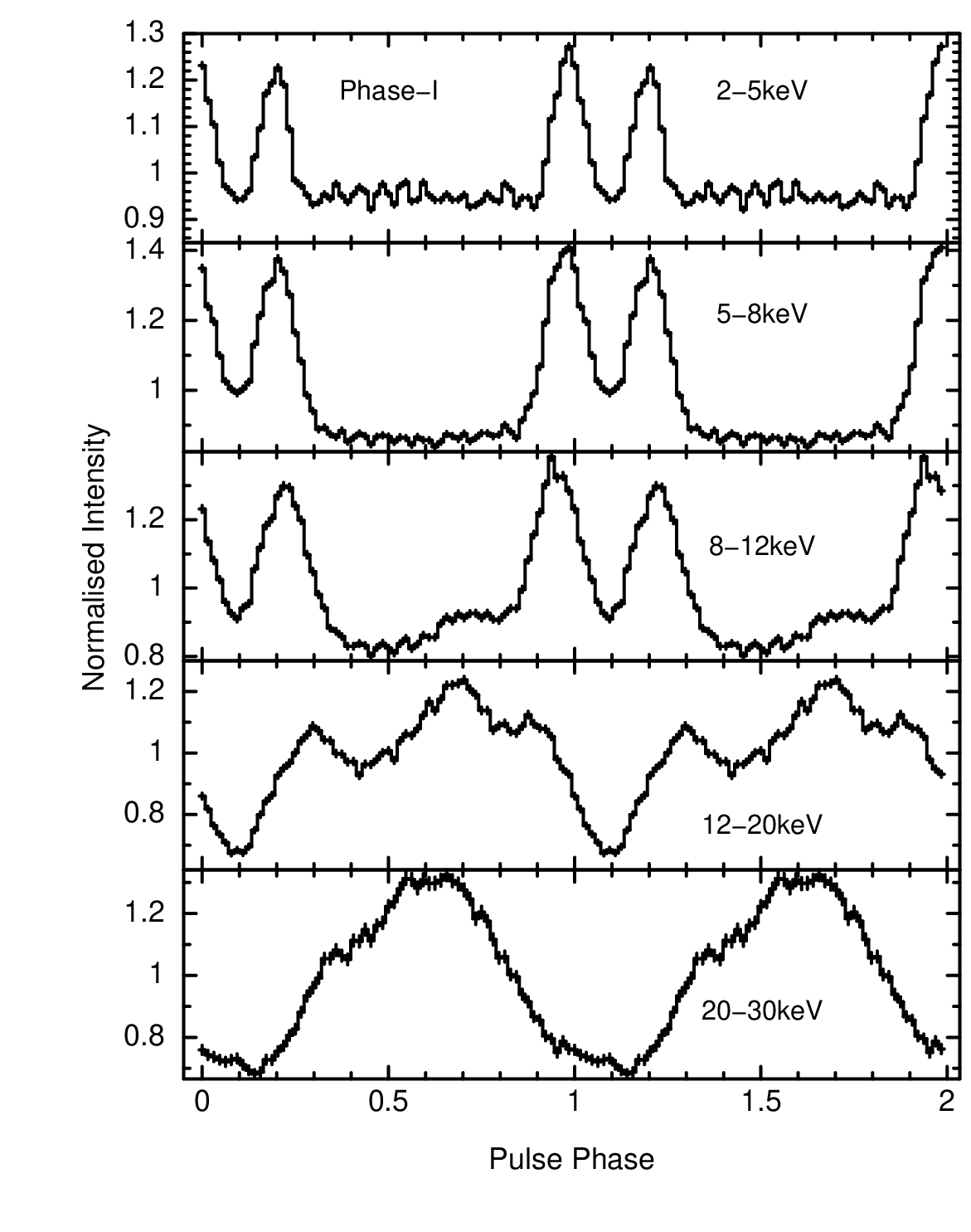}
\includegraphics[height=4.0in, width=2.0in, angle=0]{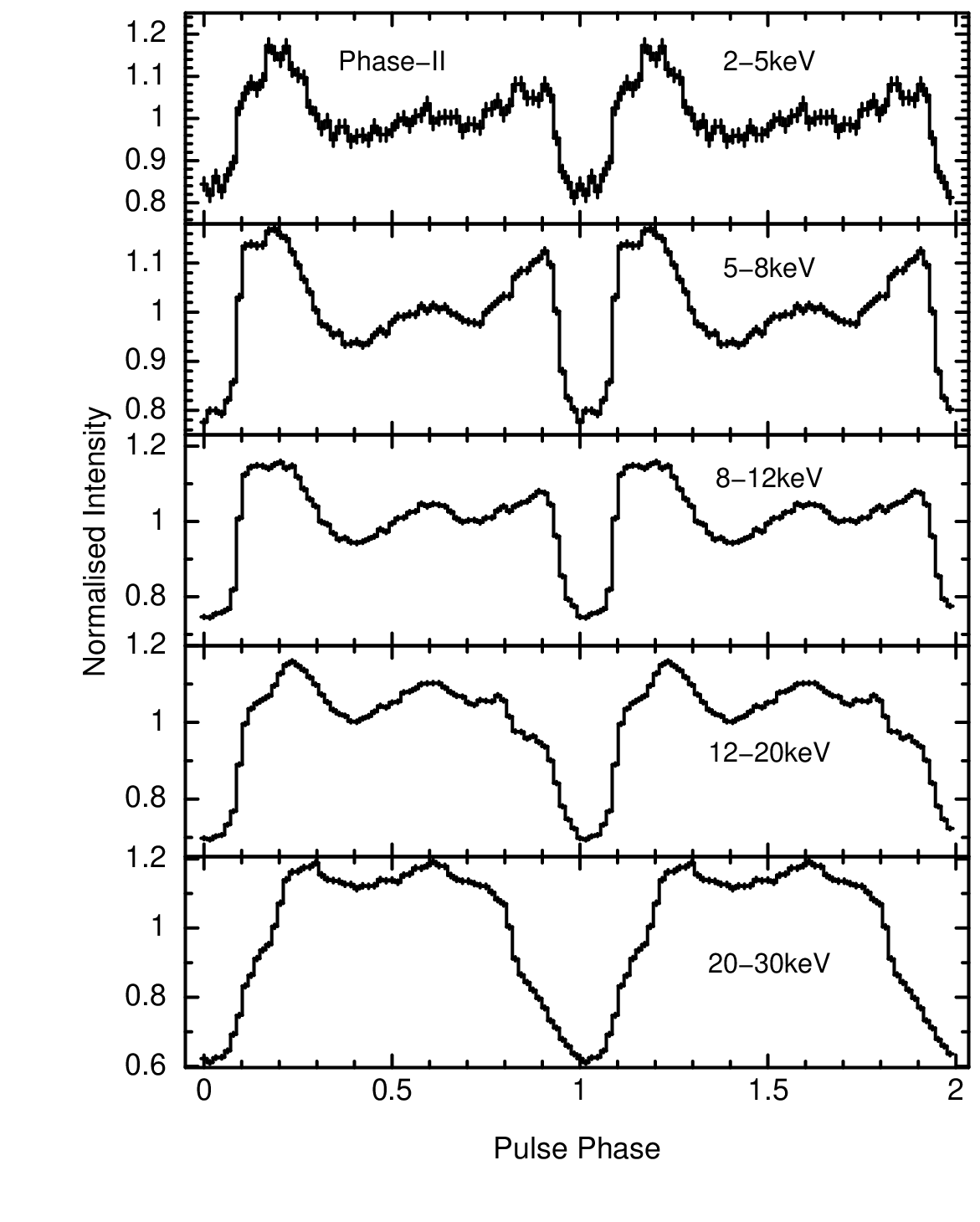}
\includegraphics[height=4.0in, width=2.0in, angle=0]{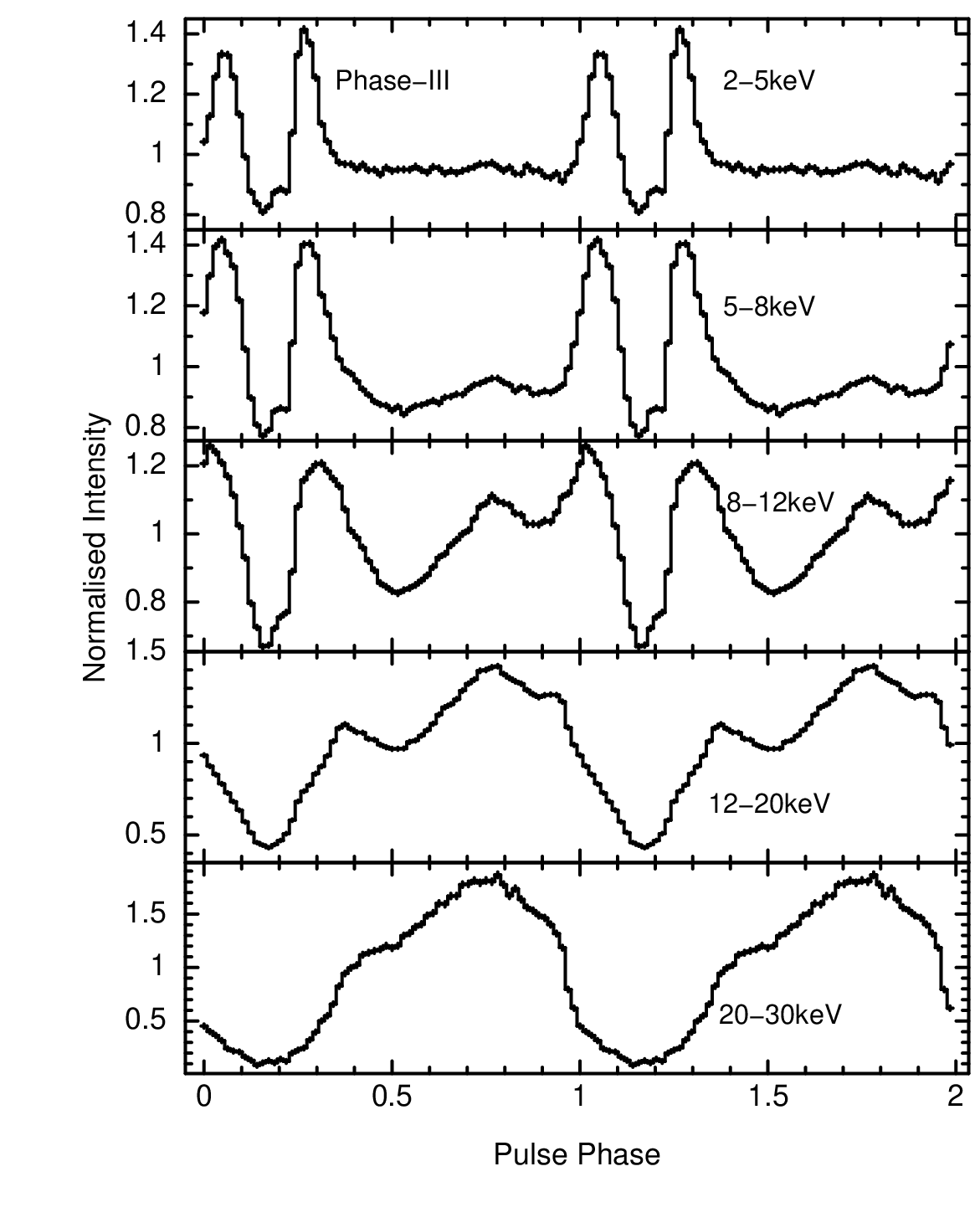}
\caption{Energy resolved pulse profile of 4U~1626--67, 
binned into 64 phasebins. Starting from the top panel, 
the energy ranges are 2-5 keV, 5-8 keV, 8-12 keV, 12-20 keV 
and 20-30 keV. The first set of panels on the left are from 
\emph{Ginga}-observations, while the other two are 
from \emph{RXTE} observations. }
\label{ene-dependent-profile}
\end{figure*}

\begin{figure*}
\centering
 \begin{minipage}{0.45\textwidth}
 \includegraphics[height=4.in,width=8.5cm]{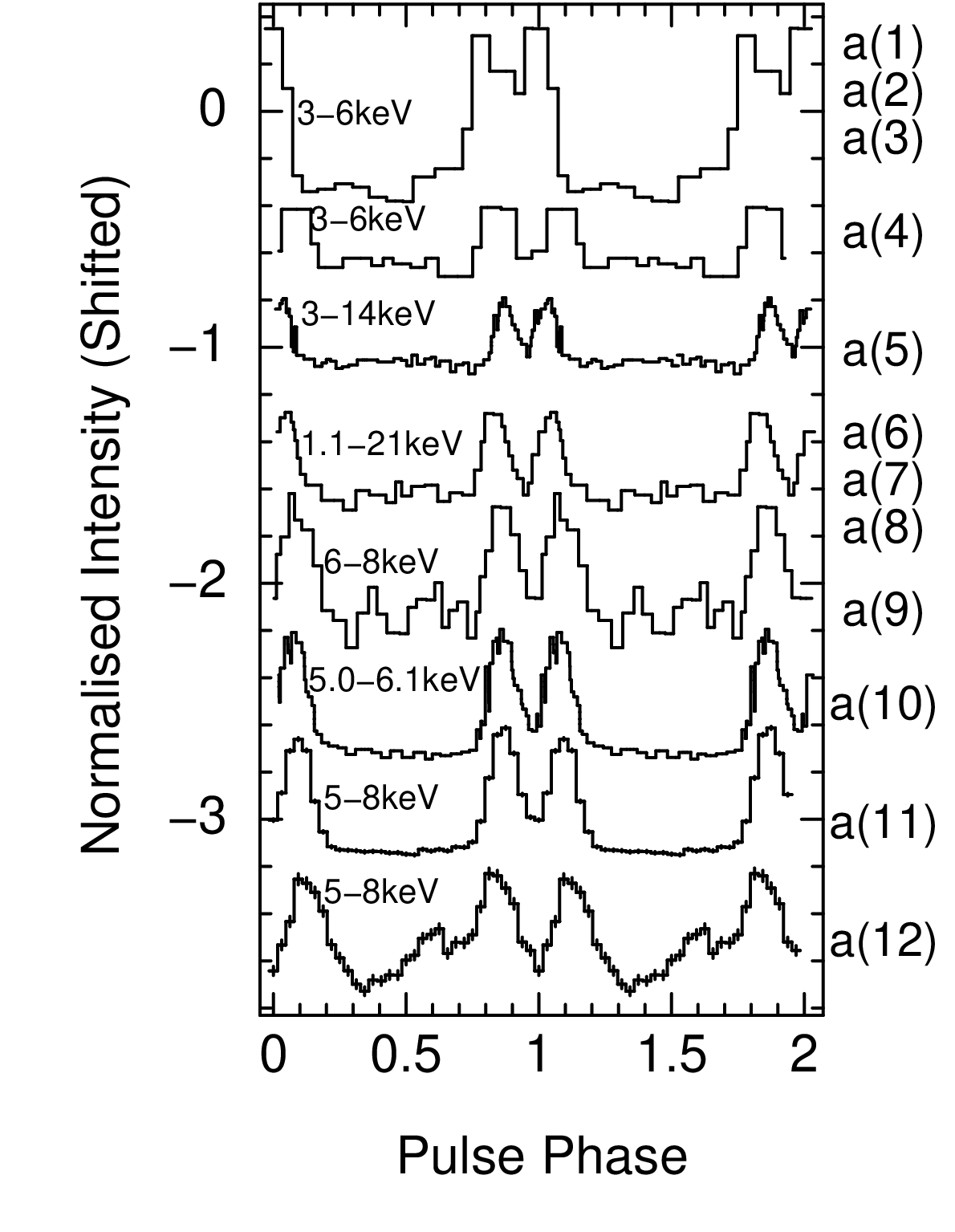}
 \includegraphics[height=4.in,width=8.5cm]{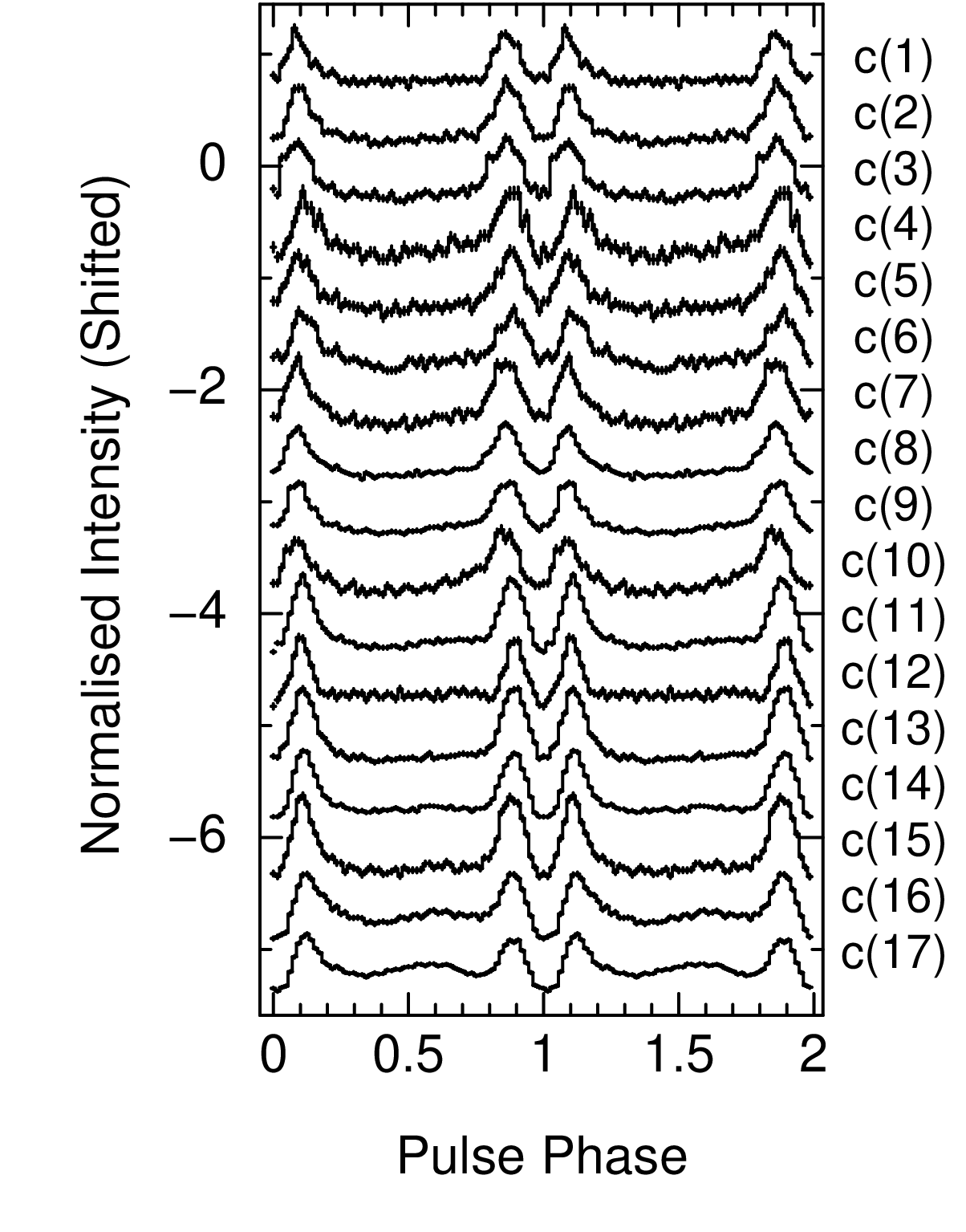}
   \end{minipage}\hfill
\begin{minipage}{0.45\textwidth}
        \includegraphics[height=8.0in,width=9.5cm]{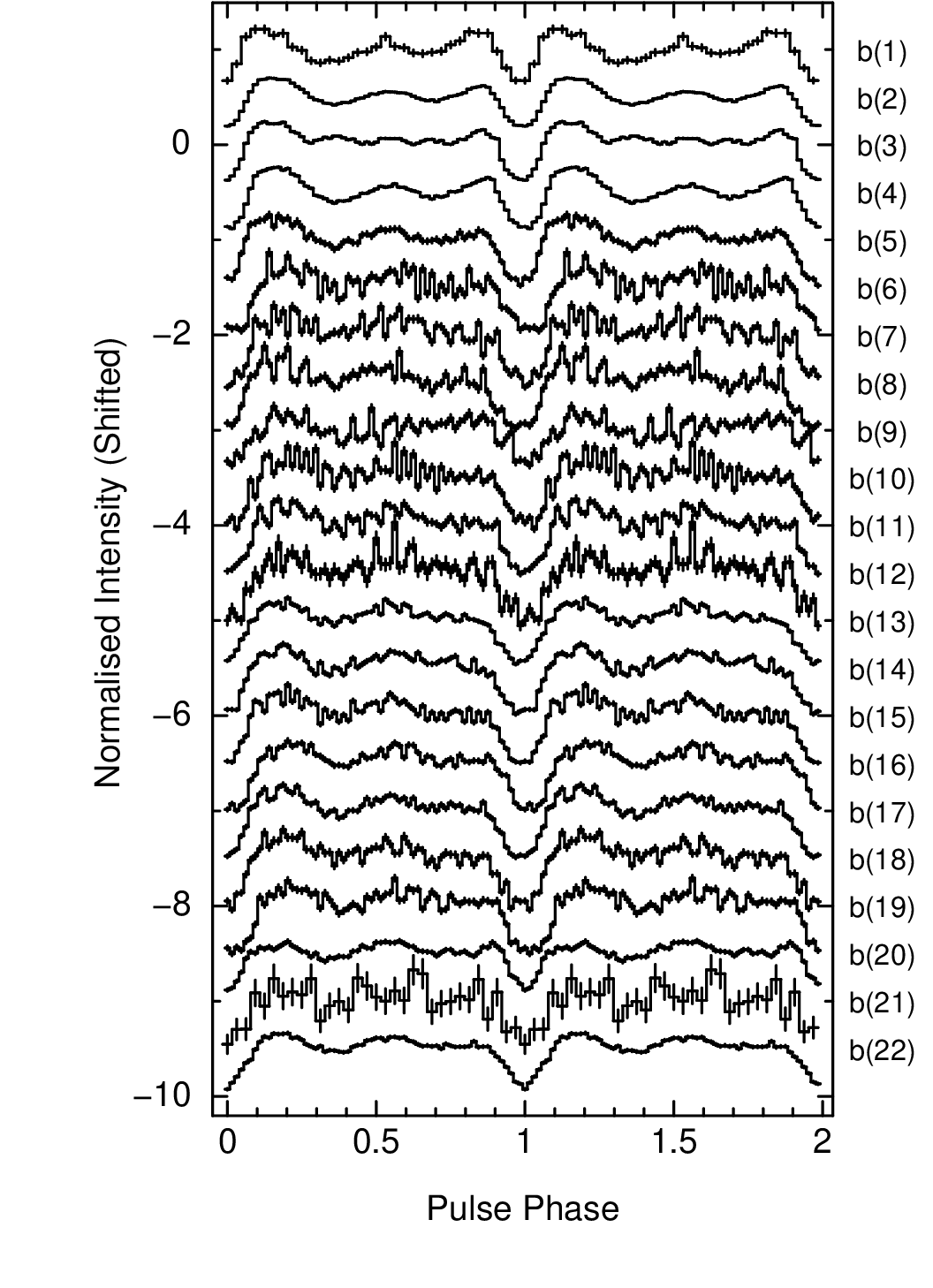}
        \end{minipage}
\caption{Pulse profiles of 4U 1626--67 using all the available observations in each of the three phases in the energy range of 5$-8$~keV, each vertically shifted for ease of viewing. The three panels clockwise from top-left are for the three phases I, II and III respectively.}
\label{profile}
\end{figure*}

\begin{table}
\centering
\caption{Observations used for the study of profile history of 4U 1626--67}
\begin{tabular}{@{}lccc@{}}

\hline
\hline

Phase & Observatory  &  Date of Observation and profile number \\
\hline
        & \emph{SAS-3}  & 1977-03-24     &        a(1)  \\
        & \emph{SAS-3}  & 1977-03-25     &        a(2)  \\
        & \emph{SAS-3}  & 1977-03-26     &        a(3)   \\
        & \emph{SAS-3}  & 1978-05-30     &        a(4)   \\
        & \emph{HEAO-I} & 1978-03-29     &        a(5)   \\
Phase I & \emph{HEAO-2}& 1979-02-24      &        a(6)   \\
        & \emph{HEAO-2}& 1979-03-10      &        a(7)   \\
        & \emph{HEAO-2}& 1979-03-14      &        a(8)   \\
        & \emph{Tenma} & 1983-05-03      &        a(9)   \\
        & \emph{EXOSAT} & 1986-03-30     &        a(10)   \\
        & \emph{Ginga}  & 1988-07-29     &        a(11)   \\
        & \emph{Ginga}  & 1990-04-19     &        a(12)    \\
\hline 

        & \emph{ASCA}-GIS & 1993-08-11   & b(1)   \\
        & \emph{RXTE}-PCA & 1996-02-10   & b(2)   \\
        & \emph{RXTE}-PCA & 1996-02-13   & b(3)   \\
        & \emph{RXTE}-PCA & 1996-02-14   & b(4)   \\
        & \emph{BeppoSAX}-MECS & 1996-08-09 &  b(5) \\
        & \emph{RXTE}-PCA & 1996-12-25   & b(6)\\
        & \emph{RXTE}-PCA & 1997-05-11   & b(7)\\
        & \emph{RXTE}-PCA & 1997-06-10   & b(8)\\
        & \emph{RXTE}-PCA & 1997-07-17   & b(9)\\
        & \emph{RXTE}-PCA & 1997-11-17   & b(10)\\
Phase II& \emph{RXTE}-PCA & 1997-12-13   & b(11)\\
        & \emph{RXTE}-PCA & 1998-06-25   & b(12)\\
        & \emph{RXTE}-PCA & 1998-06-27   &  b(13)\\
        & \emph{RXTE}-PCA & 1998-06-27  &  b(14)\\
        & \emph{RXTE}-PCA & 1998-06-28  &  b(15)\\
        & \emph{RXTE}-PCA & 1998-06-29  & b(16)\\
        & \emph{RXTE}-PCA & 1998-07-26  &  b(17)\\
        & \emph{RXTE}-PCA & 1998-07-27  & b(18)\\
        & \emph{RXTE}-PCA & 1998-07-27  &  b(19)\\
        & \emph{XMM}-PN   & 2003-08-20  &  b(20)\\
        & \emph{Swift}-XRT & 2006-02-15 &  b(21)\\
        & \emph{Suzaku}-XIS & 2006-03-09 &  b(22)\\
\hline
        & \emph{RXTE}-PCA & 2009-06-02  &  c(1)\\
        & \emph{RXTE}-PCA & 2009-06-02  &  c(2)\\
        & \emph{RXTE}-PCA & 2009-06-03  &  c(3)\\
        & \emph{RXTE}-PCA & 2010-01-14  &  c(4)\\
        & \emph{RXTE}-PCA & 2010-01-14  &  c(5)\\
        & \emph{RXTE}-PCA & 2010-01-14  &  c(6)\\
        & \emph{RXTE}-PCA & 2010-01-14  &  c(7)\\
        & \emph{RXTE}-PCA & 2010-01-14  &  c(8)\\        
        & \emph{RXTE}-PCA & 2010-01-15  &  c(9)\\
 Phase III & \emph{RXTE}-PCA &2010-01-15 &  c(10)\\
        & \emph{RXTE}-PCA & 2010-12-04  &  c(11)\\
        & \emph{RXTE}-PCA & 2010-12-04 &   c(12)\\
        & \emph{RXTE}-PCA & 2010-12-04 &   c(13)\\
        & \emph{RXTE}-PCA & 2010-12-06 &   c(14)\\
        & \emph{RXTE}-PCA & 2010-12-06 &   c(15)\\
        & \emph{RXTE}-PCA & 2010-12-07  &  c(16)\\
        & \emph{RXTE}-PCA & 2010-12-09 &  c(17)\\

\hline
\end{tabular}
\end{table}

\subsection{Pulse profile history}

We further examined similarity between the pulse profiles during the two spin-up eras and differences with the profiles in the spin-down era using all archival data of 4U 1626-67 and the published pulse profiles where archival data or necessary software is not available. We have analysed archival data and created pulse profiles from all the available observations of \emph{RXTE}-PCA during \emph{Phase II} and  \emph{Phase III} with exposure time greater than 1000 seconds. For \emph{Phase II}, we have also  analysed data from \emph{ASCA}-GIS \citep{Makishima96}, \emph{BeppoSAX}-MECS \citep{Boella95}, \emph{XMM-Newton}-PN \citep{Struder01}, \emph{Swift}-XRT \citep{Burrows05} and \emph{Suzaku}-XIS \citep{Ozawa08}. Pulse profiles in the energy range of 5-8keV were created as the characterstic features were  prominent and clear in this energy band for most of the observatories. Standard methods, appropriate for the respective instruments were used for extraction of light curves, estimation and subtraction of background, barycentre correction, pulse period determination and creation of the pulse profile in the 5-8 keV energy band. For the sake of brevity, we do not mention all the details here. For \emph{Phase I} we have analysed a \emph{Ginga}-LAC observation carried out on 29 July 1988, and for observations with \emph{SAS-3}, \emph{HEAO-1}, \emph{HEAO-2}, \emph{TENMA} and \emph{EXOSAT}-ME, we replotted the published pulse profiles (see references at the end of section 2.1.3) using the tool \emph{Dexter}\footnote{(http://dc.zah.uni-heidelberg.de/sdexter)}.

The pulse profiles of the three spin-up and spin-down eras are shown in Figure \ref{profile} in chronological order. Details of all the observations used for the same are mentioned in Table-2. Since most of the  \emph{Phase I} pulse profiles are reproduced from published articles, they are not exactly in the energy range of 5-8 keV. The energy ranges are mentioned in the figure. Within each panel, the pulse phases were aligned to have the minima at the same phase.
This figure clearly shows that:
\begin{itemize}
\item
During each of these phases, some lasting more than a decade, the pulse profile in a given energy band remains almost identical.
\item
The pulse profile in \emph{Phase I} and \emph{Phase III} are quite similar and it is very different in \emph{Phase II}.
\item
The \emph{Ginga} observation made on 19 April 1990 shows that evolution of the pulse profile started around the period when the spin-down transition began to occur.
\end{itemize}

\section{Results and Discussion}

In the present work, we have established a clear correlation between the accretion torque acting on the pulsar 4U~1626-67 and its pulse profile in the three phases of accretion torque.  \emph{Phase I} pulse profiles are double peaked and this nature of the pulse profiles disappears in \emph{Phase II}, and again reappears in \emph{Phase III}.

In the standard model of accretion onto magnetised compact stars \citep{Ghosh79}, a clear and strong correlation is expected between the mass accretion rate and the accretion torque. A positive correlation between the observed X-ray luminosity and the rate of change of spin period is found in several transient pulsars indicating that, at least in those sources, the observed X-ray luminosity is a measure of the mass accretion rate \citep{Bildsten97}. At the same time, most of the persistent pulsars do not show a clear correlation between the two quantities. In the case of 4U~1626--67, two abrupt torque reversals between spin-up and spin-down have been detected, associated with significant spectral transition. Apart from the change in the absorption column density, the spectral parameters of the two spin-up phases \emph{Phase-I} and \emph{Phase-III} are similar whereas in \emph{Phase-II}, the energy spectrum became relatively harder and the associated blackbody temperature also decreased \citep{Pravdo79,Angelini95,Kii86,Jain10,Camero12}.  However, during the first torque reversal (from spin-up to spin-down), the X-ray flux had been decreasing monotonically \citep{Angelini95, Owens97, Krauss07}. This decrease is insignificant on short timescales near the reversal and hence insufficient to explain the sudden change of accretion torque. The second torque reversal (from spin-down to spin-up) was also abrupt, but it was accompanied by an increase in the X-ray flux \citep{Camero10, Jain10}. During the entire \emph{Phase II}, the X-ray flux decreased which was associated with only a small change in the spin-down rate \citep{Krauss07}. Thus a detailed comparison of the long term X-ray flux history and the spin history, implies that if the X-ray luminosity is proportional to the mass accretion rate, then mass accretion rate is not the sole criteria affecting the transfer of torque in this system.

In addition to the shape of the energy dependent pulse profile, we have found 
several other similarities of the source character in the two spin-up phases.
Strong flares are seen only in \emph{Phase-I} and \emph{Phase-III} and not in 
\emph{Phase-II}. The PDS has a power-law shape and the QPOs are absent/weak in 
\emph{Phase-I} and \emph{Phase-III} which is contrary to its character in 
\emph{Phase-II} \citep{Jain10}.

Based on the changes in the pulse profile, \cite{Krauss07} had earlier proposed
a change in the accretion mode during the first torque reversal.
\cite{Jain10} had proposed a change in the accretion mode during the
second torque reversal mainly based on the change in the shape of the
PDS and disappearance of the QPOs in \emph{Phase III}.

Energy dependent dips are known in many X-ray pulsars. Dip features that
are prominent in the low energy band and disappear above about 7-15 keV
are explained to be due to absorption in the phase locked accretion streams
\citep{Devasia11, Maitra12}.
The low energy pulse profiles of 4U~1626--67 can also be explained in a
similar way
if there is a strong, narrow, soft emission component from the
polar hot spot, the peak of which is strongly absorbed by the same accretion
stream that produces the hot spot, giving rise to the dip feature.
In the \emph{Phases I \& III}, the absorbing structure is 
narrower than the emission beam resulting in two peaks while in \emph{Phase II}, 
the accretion stream absorbs most of the hot-spot emission. It is possible that there is another
emission component that has a broad pulse profile and dominates in the high
energy band, $>$ 20 keV. The high energy component does not vary very much
in the different phases, but the strength of the emission from the hot
spot and the strength of the absorption by the accretion stream varies,
the absorption being stronger in \emph{Phase II}.\\

The above scenario requires the size and shape of the polar hot-spot regions and the
geometry of the accretion stream to be different in the different \emph{phases} of 4U~1626--67.
Accretion from the inner disk to the neutron star surface probably happens in different
modes that are steady over many years. Transitions between such stable accretion states
with different geometry of the accretion flow can give rise to different X-ray spectra and
pulse profiles as observed in 4U~1626--67. An interesting possibility is a radiation pressure
induced warping of the inner accretion disk which may become retrograde leading to negative
accretion torque of similar magnitude \citep{Kerkwijk98}. The same authors have also
pointed out that what sets the time scale for such transitions in different sources is not
clear. While the QPO features may be different in different accretion states, it is also to
be noted that there are fewer flares in the spin-down era. 
\\

 A change in accretion mode in \emph{Phase-II} compared to the other two phases can also result into the observed energy dependent pulse profiles in the following manner. 
In the two spin-up eras there may be
a steady accretion column formed which feeds the neutron star at a constant mass 
accretion rate. This accretion column intercepts the emission cone such as to form
the narrow dips seen between the double peaks. During \emph{Phase II} the accretion
is probably from a more clumpy material roughly forming an accretion column which
manifests itself with pulse phase dependent variation in the absorption column density 
explaining the increase in the number of features in the average pulse profile. This
may also explain the presence of QPOs during \emph{Phase II} if we assume that QPOs
are a manifestation of clumps in the inner accretion disk.

We also note here some other possibilies that were pointed out earlier to
explain the torque reversals in 4U~1626--67 and some associated characteristics.
\cite{Nelson97} had proposed that a transition between prograde-retrograde
disk causes the torque reversal in 4U~1626--67. While formation of a
retrograde disk is quite possible in wind fed systems, in the case of
Roche lobe overflow systems the possibility of a retrograde disk formation and it
being present for an extended period is not certain.
Switching between Keplerian and sub-Keplerian rotation of the accretion flow
\citep{Narayan95, Yi99} was also proposed to be a reason for the change of
state. In the present work we have given strong observational evidences for
a change in the accretion mode during the torque reversals, which is
possibly in the form of changes in the accretion flow geometry from the
inner part of the accretion disk to the neutron star.

\section*{Acknowledgments}
We would like to thank all the members of \emph{Ginga} for their support in the software installation. This research has made use of data obtained from the Data ARchives and Transmission System (DARTS) and High Energy Astrophysics Science Archive Research Center (HEASARC), provided by NASA’s Goddard Space Flight Center.  \\

\label{lastpage}

\begin{thebibliography}{}
\bibitem[\protect\citeauthoryear{Angelini et al.}{1995}]{Angelini95}
Angelini L., et al., 1995, ApJ, 449, L41
\bibitem[\protect\citeauthoryear{Bildsten et al.}{1997}]{Bildsten97}
Bildsten L., Chakrabarty D., Chiu J., Finger M. H., Koh D. T., Nelson R. W., Prince T. A., Rubin B. C., Scott D. M., Stollberg M., Vaughan B. A., Wilson C. A., Wilson R. B., 1997, ApJS, 113, 367
\bibitem[\protect\citeauthoryear{Bradt, Rothschild \& Swank}{Bradt et al.}{1993}]{Bradt93}
Bradt H. V., Rothschild R. E., Swank, J. H., 1993, A\&AS, 97, 1
\bibitem[\protect\citeauthoryear{Boella et al.}{1995}]{Boella95}
Boella G., BeppoSAX team, 1995, \textit{Proc. SPIE Conference}, 2517-14
\bibitem[\protect\citeauthoryear{Burrows  et al.}{2005}]{Burrows05}
Burrows David N., et al., 2005, \textit{Space Science Reviews}, 120, 3-4,  165-195
\bibitem[\protect\citeauthoryear{Coburn et al.}{2002}]{Coburn02}
Coburn, W., Heindl, W. A., Rothschild, R. E., Gruber, D.E., Kreykenbohm, I., Wilms,J., Krteschmar,P., Staubert,R. 2002, APJ, 580,394
\bibitem[\protect\citeauthoryear{Camero-Arranz et al.}{2010}]{Camero10}
Camero-Arranz A., Finger M. H., Ikhsanov N. R., Wilson-Hodge C. A., Beklen E., 2010, ApJ, 708, 1500
\bibitem[\protect\citeauthoryear{Camero-Arranz et al.}{2012}]{Camero12}
Camero-Arranz, A., Pottschmidt, K., Finger, M. H., Ikhsanov, N. R.,
Wilson-Hodge, C. A., Marcu, D. M., 2012, A\&A, 546, 40
\bibitem[\protect\citeauthoryear{Chakrabarty et al.}{1997}]{Chakrabarty97}
Chakrabarty D., et al., 1997, ApJ, 474, 414
\bibitem[\protect\citeauthoryear{Chakrabarty et al.}{1998}]{Chakrabarty98}
Chakrabarty D., 1998, ApJ, 492, 342
\bibitem[\protect\citeauthoryear{Devasia et al.}{2011}]{Devasia11}
Devasia, J., James, M., Paul, B., Indulekha, K., 2011, MNRAS, 417, 348
\bibitem[\protect\citeauthoryear{Elsner et al.}{1980}]{Elsner83}
Elsner R. F., Ghosh P., Lamb F. K., 1980, Ap
J, 241, L155
\bibitem[\protect\citeauthoryear{Ghosh \& Lamb}{1979}]{Ghosh79}
Ghosh P., Lamb F. K., 1979, ApJ, 234, 296
\bibitem[\protect\citeauthoryear{Giacconi et al.}{1972}]{Giacconi72}
Giacconi R., Murray S., Gursky H., Kellogg E., Schreier E., Tananbaum H., 1972, ApJ, 178, 281
\bibitem[\protect\citeauthoryear{Heindl et al.}{1999}]{Heindl99}
Heindl, W. A., \& Chakrabarty, D. 1999, in Highlights in X-ray Astronomy,
ed. B. Aschenbach \& M. J. Freyberg (MPE Rep. 272; Garching: MPE), 25
\bibitem[\protect\citeauthoryear{Iwakiri et al.}{2012}]{Iwakiri12}
Iwakiri W. B., Terada Y., Mihara T., Angelini L., Tashiro M. S., Enoto T., Yamada S., Makishima K., Nakajima M., Yoshida A., 2012, APJ, 751, 35
\bibitem[\protect\citeauthoryear{Jahoda et al.}{1996}]{Jahoda96}
Jahoda K., Swank J. H., Giles A. B., Stark M. J., Strohmayer T., Zhang W., Morgan E. H., 1996, SPIE, 2808, 59
\bibitem[\protect\citeauthoryear{Jahoda et al.}{2006}]{Jahoda06}
Jahoda K., Markwardt C. B., Radeva Y., Rots A. H., Stark M. J., Swank J. H., Strohmayer T. E., Zhang W., 2006, ApJS, 163, 401
\bibitem[\protect\citeauthoryear{Jain, Paul \& Dutta}{2008}]{Jain08}
Jain C., Paul B., Dutta A., Joshi K., Raichur H., 2008, JApA, 28, 175J	
\bibitem[\protect\citeauthoryear{Jain, Paul \& Dutta}{2010}]{Jain10}
Jain C., Paul B., Dutta A., 2010, MNRAS, 403, 920J
\bibitem[\protect\citeauthoryear{Joss, Avni \& Rappaport}{1978}]{Joss78}
Joss P. C., Avni Y., Rappaport S., 1978, ApJ, 221, 645
\bibitem[\protect\citeauthoryear{Kaur, Paul \& Sagar}{2008}]{Kaur08}
Kaur R., Paul B., Sagar R., 2008, ApJ, 676, 1184
\bibitem[\protect\citeauthoryear{Kerkwijk et al.}{1998}]{Kerkwijk98}
Kerkwijk van M. H., Chakrabarty D., Pringle J. E., \& Wijers R. A. M. J., 1998, APJ, 499, L27–L30
\bibitem[\protect\citeauthoryear{Kii et al.}{1986}]{Kii86}
Kii T., Hayakawa S., Nagase F., Ikegami T., Kawai N., 1986, PASJ, 38, 751
\bibitem[\protect\citeauthoryear{Kommers, Chakrabarty \& Lewin}{1998}]{Kommers98}
Kommers J. M., Chakrabarty D., Lewin W. H. G., 1998, ApJ, 497, L33
\bibitem[\protect\citeauthoryear{Krauss et al.}{2007}]{Krauss07}
Krauss M. I., Schulz N. S., Chakrabarty D., Juett A. M., Cottam J., 2007, ApJ, 660, 605
\bibitem[\protect\citeauthoryear{Li et al.}{1988}]{Li88}
Li, F. K., McClintock, J., Rappaport, S., van der Klis, M., \& Verbunt, F. 1988, ApJ, 327, 732
\bibitem[\protect\citeauthoryear{Li et al.}{1980}]{Li80}
Li, F. K., McClintock, J. E., Rappaport, S., Wright, E. L., \& Joss, P. C. 1980, ApJ, 240, 628
\bibitem[\protect\citeauthoryear{Levine et al.}{1988}]{Levine88}
Levine A., Ma C. P., McClintock J., Rappaport S., van der Klis M., Verbunt F. 1988, ApJ, 327, 732
\bibitem[\protect\citeauthoryear{Makino \& Astro-C team}{1987}]{Makino87}
Makino F., Astro-C team, 1987, \textit{Atrophys. Letters Commun.}, 25, 223
\bibitem[\protect\citeauthoryear{Makishima \& Astro-D team}{1996}]{Makishima96}
Makishima, K., Astro-D team, 1996, \textit{PASJ}, 48, 171
\bibitem[\protect\citeauthoryear{Maitra, Paul \& Naik}{2012}]{Maitra12}
Maitra, C., Paul, B., Naik, S., 2012, MNRAS, 420, 2307
\bibitem[\protect\citeauthoryear{McClintock et al.}{1980}]{McClintock80}
McClintock, J. E., Li, F. K., Canizares, C. R., \& Grindlay, J. E. 1980, ApJ, 235, L81
\bibitem[\protect\citeauthoryear{Narayan \& Yi}{1995}]{Narayan95}
Mihara, T., 1995, Ph.D.thesis, Univ.Tokyo
\bibitem[\protect\citeauthoryear{Mihara}{1995}]{Mihara95}
Narayan R., Yi I., 1995, ApJ, 452, 710
\bibitem[\protect\citeauthoryear{Nelson et al.}{1997}]{Nelson97}
Nelson R. W., et al., 1997, ApJ, 488, L117
\bibitem[\protect\citeauthoryear{Orlandini et al.}{1998}]{Orlandini98}
Orlandini, M., Fiume, D. Dal, Frontera, F., del Sordo, S., Piraino, S.,
Santangelo, A., Segreto, A., Oosterbroek, T., Parmar, A. N., 1998, ApJ, 500, L163
\bibitem[\protect\citeauthoryear{Ozawa et al.}{2008}]{Ozawa08}
Ozawa Midori., et al., 2008, \textit{Proc. SPIE, Space Telescopes and Instrumentation}, 70112B
\bibitem[\protect\citeauthoryear{Owens, Oosterbroek \& Parmar}{1997}]{Owens97}
Owens A., Oosterbroek T., Parmar A. N., 1997, A\&A, 324, L9
\bibitem[\protect\citeauthoryear{Pravdo et al.}{1978}]{Pravdo78}
Pravdo S. H., Bussard R. W., Becker, R. H., Boldt, E. A., Holt, S. S., and Serlemitsos, P. J. 1978, APJ, 225, 988 
\bibitem[\protect\citeauthoryear{Pravdo et al.}{1979}]{Pravdo79}
Pravdo S. H., et al., 1979, ApJ, 231, 912
\bibitem[\protect\citeauthoryear{Rappaport et al.}{1977}]{Rappaport77}
Rappaport S., Markert T., Li F. K., Clark G. W., Jernigan J. G., McClintock J. E., 1977, ApJ, 217, L29
\bibitem[\protect\citeauthoryear{Schulz et al.}{2001}]{Schulz01}
Schulz, N. S., Chakrabarty, D., Marshall, H. L., Canizares, C. R., Lee, J. C., Houck, J., 2001, ApJ, 563, 941
\bibitem[\protect\citeauthoryear{Schulz et al.}{2011}]{Schulz11}
Schulz, N. S., Marshall, H. L., Chakrabarty, D. 2011, AAS, 21812205

\bibitem[\protect\citeauthoryear{Shinoda et al.}{1990}]{Shinoda90}
Shinoda K., et al., 1990, PASJ, 42, L27
\bibitem[\protect\citeauthoryear{Vaughan \& Kitamoto}{1997}]{Vaughan97}
Vaughan B. A., Kitamoto S., 1997, preprint (astro-ph/9707105)
\bibitem[\protect\citeauthoryear{White, Swank \& Holt}{1983}]{White83}
White N. E., Swank J. H., Holt S. S., 1983, Astrophys. J., 270, 711
\bibitem[\protect\citeauthoryear{Struder et al.}{2001}]{Struder01}
Struder L., et al., 2001, A\&A 365, L18

\bibitem[\protect\citeauthoryear{Yi \& Vishniac}{1999}]{Yi99}
Yi I., \& Vishniac E.T., 1999, APJ, 516, L87
\end{thebibliography}
\end{document}